# CONFORMATIONAL TRANSFORMATIONS OF DNA MACROMOLECULE IN HETERONOMOUS CONFORMATION

**P.P. Kanevska, S.N.Volkov**
Bogolyubov Institute for Theoretical Physics, NAS of Ukraine
snvolkov@bitp.kiev.ua

To understand the mechanism of TATA-box conformational transformations we model structure mobility and find the types of conformational excitations of DNA macromolecule in heteronomous conformation. We have constructed the two-component model for describing DNA conformational transformation with simultaneous transitions in the furanos rings of the monomer link. Internal component describes the change of the base pair position in the double helix. External component describes the displacement of mass center of the monomer link. Nonlinearity of the system is accounted with a form of potential energy describing C3'→C2' and C2'→C3' sugars transitions in monomer link, and interrelation between monomer conformational transition and macromolecule deformation. The comparison of our results with experimental data [15] allows to confirm that the localized conformational excitations may realise in DNA TATA-box. These excitations cause the deformation of the macromolecule fragment.
**KEY WORDS:** DNA**,** heteronomous conformation, conformational transformations, TATA-box.

The genetic activity of the DNA essentially depends on the conformational and spatial forms, which a macromolecule can have. DNA conformation modifications define the efficiency of interactions with proteins, recognition of definite sites of the macromolecule and, in general, a regulation of genetic processes. Similarly, DNA TATA-box, a core gene promoter in eukaryotes, is recognized with a multi-protein complex through capacity of this DNA fragment to dramatically bend according to the saddle-like shape of TBP contacting with the TATA-box (TBP = TATA-binding protein). In spite of severe deformation, the TATA-box bound to TBP saves double strand homogenous form relatively phosphates.We argue that the unique change of conformation in the fragment is associated with conformational properties of alternating structure T-A-T-A [1]. So TATA-box DNA bend is a response on the specific conformation transition. Here is a short review of studies investigating alternating T-A DNA sequences.

A.Klug was the first who paid attention to a biological role of conformation properties of this polynucleotide sequence, when he studied linkages of lac-repressor protein with different DNA sequences [2]. For poly (AT) ×poly (AT) the constant of linkage appeared three orders less, than for thymus DNA. In the article [2], the selectivity of a recognition was shown to be connected with distinct conformation of bound to protein poly(AT) ×poly(AT) from the customary forms. This form was called "alternating $B$"- form by Klug. Studying the diffraction $\gamma$ - rays analysis of a the tranucleotide dA-dT-dA-dT crystal, he suggested, that the singularity of poly (AT) ×poly (AT) fine structure is conditioned by relative instability of A-T step. This means that $_{5'}T_{3'} - {}_{5'}A_{3'}$ staking is smaller than $_{5'}A_{3'} - {}_{5'}T_{3'}$ staking. Then the torsion angles $\omega'$(O-3'-P) and $\omega$ (P-O-5') for alternating strand have the value $\omega' = \omega = -60^o$ between A and T, and $\omega' = -120^o$, $\omega = -50^o$ between T and A (for the customary B - form the torsion angles equal $-90^o$ and $-60^o$, accordingly). In this case the form of sugar bound with a thymine should be C'2 endo, and bound with adenin should be C'3 endo [2]. In 1980, outcomes of $^{31}$P NMR study [3] have shown, that synthetic DNA poly (AT) ×poly (AT) in fibers has inhomogeneous (heteronomous) conformation with alternation of $A$ and $B$ forms in a frame similarly to crystal stucture. Later Raman and the infrared spectroscopy studies have confirmed coexisting of sugers in C'2 endo and C'3 endo forms in links of alternating adenin-thimine sequences of DNA [4-12].

The attention to conformation properties of alternating T-A DNA has increased after X-ray analysis of crystal configuration of TATA-box and TBP complex [13-15]. It is known that the sequence of TATA-box or TATA@A@N (where @ is T or A base, and N - any base) plays a relevant role in the regulation of a process of transcription, uniquely determining a direction and the beginning of transcription. The experiments and further simulations of molecular dynamics of a complex TATA-box+TBP have shown that the TATA-box is characterized with high conformational flexibility. Due to this flexibility a multiprotein complex recognizes TATA-box. The binding TBP to TATA-box causes the deformation of DNA: a piece of a macromolecule containing 8 pairs is bent on $80^o$ with a simultaneous unwind of a spiral in general on $110^o$. The deformation goes in such a way that the helical structure is preserved, and the macromolecule remains in $B$-conformation outside the TATA-box. Despite numerous experimental [13-15] and numerical [16-21] studies, the nature of bending mechanisms remains unclear.

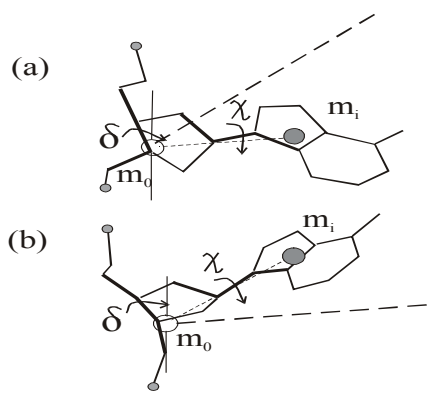
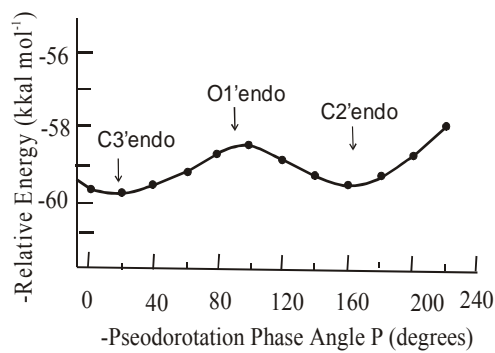

Fig.1: (a) – nucleotide form for C'$_3$ *endo* sugar; (b) – nucleotide form for C'$_2$ *endo* sugar. $\chi$ - the glycosyl torsion between sugar and base, $\delta$ - the backbone torsion angle associated with the sugar ring

Fig.2: The potential energy of nucleotide [27,28].

The purpose of this study is the modeling of structural transformations of alternating T-A sequences of the DNA in heteronomous conformation and the defining the influence of conformation excitations in these patterns on deformations of a chain as the whole. For the description of structural transformations of a double helix in heteronomous conformation we have used the approach of modeling developed by S. Volkov [22-26]. The characterizing parameters of the model correspond to the mechanical properties and structural organization of the heteronomous DNA macromolecule. In this paper we present the model of heteronomous transformation and static excitations of the system. We also determine a part of deformation caused by intrinsical conformational transformation. Quantitative estimations agree with experiment [15].

### CONSTRUCTION OF THE MODEL OF STRUCTURAL TRANSFORMATIONS OF HETERONOMOUSE DNA

The characteristics property of heteronomous conformation of DNA is a conformational irregularity along a double helix. It is possible to understand from experiments conducted in [2-12] that the irregularity of alternating polynucleotide - poly (AT) poly (AT) is exhibited in alternation of C'2 endo and C'3 endo conformations and in alternation of phosphate groups of a frame conformations responding to the forms of a sugars in *A* and *B* conformations. Experimental data are satisfied by two possible conformations of a double helix. The first possible conformation is formed in a way that double helix contains sugars in C'3 endo for tymin residue and in C'2 endo for adenin residue. The second possible conformation is realised in a way that in one chain of a double helix all sugars have C'2 endo form, and in other chain all sugars have C'3 endo form. Because of cooperativity of structural organization, the macromolecule must be zigzag form for the first type of the conformation, similar to *Z*-conformation of DNA. This feature has not been proves by experiments. While the second type of conformation must have secondary structure with the homogeneous backbone along a chain. The last one is more likely to be realised in vivo. Besides, the analysis of phosphate groups' disposition in [15] demonstrates that conformation with homogeneous backbone is realized in the DNA TATA-box. Therefore, not disclaiming the possibility of the first type conformation, we shall consider structural transformations within the framework of a homogeneous frame. The term "heteronomous conformation" will be used when talking about the second type.

The DNA double helix in heteronomous conformation is modelled as a chain of monomers. Each monomer consists of a pair complementary nucleotides. The nucleotide is considered as two masses: one is formed by atoms of sugar-phosphate backbone, and other - by the base with sugar. Thus monomer consists of 4 masses. Despite the simplicity of this approach, it works in the case of small amplitude oscillations and for obtaining of all dominant modes of low frequency spectra A- and B-forms of the DNA [22-24]. Starting from the model of 4-masses, it has been possible to model such structural transformations as B-A transition and preopening of pairs. The structural transformations have been described in terms of coordinates of groups of structure elements moving in coordination [25,26]. In this article we apply this approach to the modelling of structural transformations in heteronomous DNA.

According to the approach [22-26], each monomer in heteronomous chain consists of four atomic groups: two masses for sugar-phosphate backbone $m_0$ and two masses for bases conjunct with sugars $m_1$ and $m_2$. Evidently, the conformations differ in a relative position of the four monomers masses. Let us define the structural parameters of a monomer link for description transformations in heteronomous DNA within the framework of model of four masses.

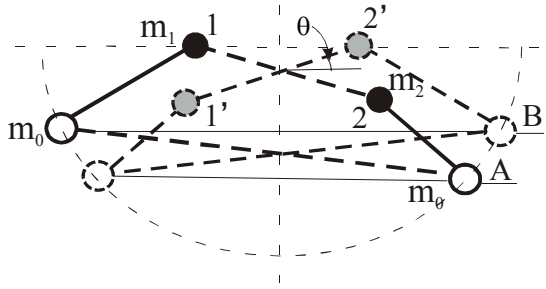 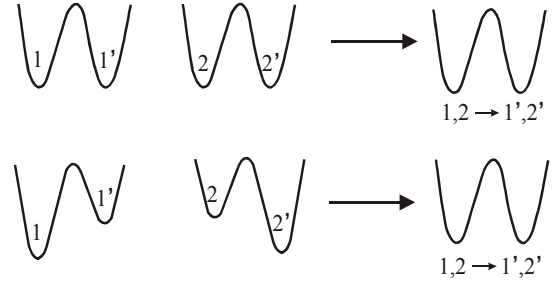

Fig.3: Two monomer states of heteronomous DNA         Fig.4: Constructing of the potential energy for monomer link.

Pseudorotation phase angle P determining the form of sugar for nucleic acid is known to have two minima. The first is in C'2 endo form of sugar and the second is in C'3 endo form [27,28] (fig. 2). In order to describe the conformations of DNA fragments, the pair of torsion angles: glycoside junction between sugar and base $\chi$ (O4'-C1'-N9/N1-C1/C2) and the backbone torsion angle associated with the sugar ring $\delta$ (C5'-C4'-C3'-O3') (fig. 1) are usually used. The form of a nucleotide is, therefore, determined by a sugars form and a nucleoside position relatively a backbone. In the two dimensional phase space of noted parameters A and B forms of double helix occupy different or weakly overlapping regions [29]. Also a pair of nucleotides in the A- form is known to be shifted along the dyadic axis of the pair in relation to the position of a pair in the B-form (fig.3).

For heteronomous DNA one of nucleotides of monomer link is described with parameters that are typical for B-form. The other nucleotide has parameters characterizing A-form. According noted features of B- and A-form, within the framework of four masses model, the difference between A- and B-form of a monomer is determined by the distance between $m_0$ and $m_i$ ( it is smaller approximately on 1 Å in the A-form, than in the B-form), and by the torsion angle of nucleoside relatively axis joining phosphate masses in a monomer (it is smaller approximately on 5° for the A-form, than for the B-form) [22]. Therefore, the monomer of the DNA macromolecule in the heteronomous conformation with a homogeneous frame can be imagined in the following way. For example, $m_1$ is in the position 1 with the sugar form C'$_2$ endo, and $m_2$ is in the position 2 with sugar form C'$_3$ endo (fig. 3). Transition from the state 1,2 in the state1', 2' means that displacements of mass $m_1$ from the position 1 in 1', and of mass $m_2$ from a position 2 in 2' occur simultaneously. Then, the potential function determined with a sugar form has a double well for a monomer as the whole as for each nucleotide. But the potential function of a monomer as a whole has the symmetrical form for both condition of A-B-equilibrium and condition of the metastablility of one of the conformations (fig. 4). The barrier of transition has an order of 3kkal/mol [31,32] that is typical for B-A transformations of a double helix of DNA.

According to the model of 4-masses, the energy of a monomer chain can be written as:

$$E = \frac{1}{2}\sum_{n=1}^{N}\sum_{i=1,2}\left\{m_0\dot{\vec{R}}_i^2 + m_i\dot{\vec{r}}_i^2 + U(\vec{R}_i(n),\vec{r}_i(n))\right\}, \qquad (1)$$

where $\vec{R}_i(n)$ is the displacement radius-vector of $m_0$ mass corresponding to i-base of pair in n-monomer, and $\vec{r}_i(n)$ is the displacement radius-vector of $m_i$ mass. The potential energy U consists of three terms:

$$U = U_1(n) + U_2(n,n-1) + U_3(n,n\pm 1). \qquad (2)$$

Here $U_1$ is the energy of conformational mobility of a monomer link structure elements, $U_2$ is the interaction energy of the nearest neighbours, $U_3$ - the energy of components correlation of the system that depends on a pathway of structural elements joint motions.

The variables in the equation (1) are vectors and, therefore, a Lagrange equation is very complicated. Therefore, in order to describe conformational properties of the system we change over to scalar variables describing joint motions of the structural elements according to pathways of transition $1,2 \rightarrow 1',2'$. To derive joint motions of monomer members corresponding to the transition, we introduce a displacement of a sugar-phosphate backbone center of mass and a center of mass of a base pair for a monomer:

$$\vec{r}_p = \frac{m_1\vec{r}_1 + m_2\vec{r}_2}{m_p}, \quad m_p = m_1 + m_2; \qquad (3)$$

$$\vec{\delta} = \vec{r}_1 - \vec{r}_2, \quad \mu_p = \frac{m_1 m_2}{m_p}; \tag{4}$$

$$\vec{r}_q = \frac{m_0 \vec{R}_1 + m_0 \vec{R}_2}{m_q}, \quad m_q = 2m_0; \tag{5}$$

$$\vec{\Delta} = \vec{R}_1 - \vec{R}_2, \quad \mu_q = \frac{1}{2} m_0, \tag{6}$$

where $r_p$ and $r_q$ describe a center of mass of a nucleosides pair and a pair of phosphates in a monomer, $m_p$ and $m_q$ are corresponding masses; $\delta$ and $\Delta$ are relative displacement of nucleosides and relative displacement of phosphates in a monomer, $\mu_p$ and $\mu_q$ are corresponding reduced masses. Then displacement of monomer center of mass $R$ and relative displacement of the nucleoside center of mass relative a center of mass of phosphates $\nu$ is:

$$\vec{R} = \frac{m_p \vec{r}_p + m_q \vec{r}_q}{M}, \quad M = m_p + m_q; \tag{7}$$

$$\vec{v} = \vec{r}_p - \vec{r}_q, \quad m = \frac{m_p m_q}{M}, \tag{8}$$

where $M$ is the mass of a monomer, and $m$ is the reduced mass. Then the kinetic energy can be written as:

$$K = \frac{1}{2} \sum_n \left[ M \dot{\vec{R}}_n^2 + m \dot{\vec{v}}_n^2 + \mu_p \dot{\vec{\delta}}_n^2 + \mu_q \dot{\vec{\Delta}}_n^2 \right]. \tag{9}$$

Because the conformational transition in an alternate state goes with syn-bate increasing and decreasing of lengths of the corresponding nucleosides, the distance between their centers of masses almost does not change. Also the distance between nucleotides centers of masses remains almost immovable relative to a phosphates center of mass. Practically, the pair of nucleosides rotates around its center of mass. So, $\dot{\vec{v}} = 0$, and instead of $\dot{\vec{\delta}}$ it is possible to enter variable $\dot{\sigma} = |\vec{\delta}| \dot{\theta}$, where $\theta$ is a turn angle of a pair of nucleosides (fig.3), $\sigma$ is the scalar distance between 1,2 and 1',2' states for centers of mass of nucleosides. According to the estimated parameters of B-A transformations, $\sigma$ changes from $-a$ to $a$ ($|a| \approx 2\text{Å}$). Also, it is important to note that the potential energy of interaction between phosphates is much larger along the chain than it is in a monomer. Therefore, a contribution from $\vec{\Delta}$ in the potential energy can be neglected and $\vec{\Delta}$ - coordinate becomes cyclic.

During B-A transition in DNA, a center of mass of monomer link shifts and chain becomes deformed [25-26]. As a rule, deformation is realized by a bend and torsion of macromolecule. The bend and the torsion correlate with each other. To simplify the problem, we can take into account one of the macromolecular chain deformational degrees of freedom. Let us consider the bend as a response to conformational transformations. Therefore, monomer displacement can be written in a scalar form: $\vec{R} \to R$.

Taking into account presented speculations, kinetic energy yields:

$$K = \frac{1}{2} \sum_n \left[ \mu_p \dot{\sigma}_n^2 + M \dot{R}_n^2 \right]. \tag{10}$$

Thus, the problem is reduced to a two components model, where $\sigma_n$ describes movability of the nucleosides pair in a n-monomer. This is an internal component. $R_n$ describes movability of monomer as the whole. This is external component.

The total potential energy is:

$$U = \sum_n \Phi(\sigma_n) + \sum_n \left[ k(R_n - R_{n-1})^2 + g(\sigma_n - \sigma_{n-1})^2 \right] + \sum_n \chi(R_{n+1} - R_{n-1}) F(\sigma_n), \tag{11}$$

where $\chi, g, k$ are parameter of components correlation and force constants of external and internal subsystems, respectively.

The first term of expression (11) describes the energy of a conformational transition. As we noted above, the monomer state can be described by a two-well potential function:

$$\Phi(\sigma_n) = \varepsilon\left[\left(\frac{\sigma_n}{a}\right)^2 - 1\right]^2. \tag{12}$$

Here $\varepsilon$ is a transition barrier between alternative conformations.

The second term of potential energy (11) describes interaction along the chain in the nearest neighbours approximation.

The third term describes correlation of internal and external components. The view of $F(\sigma_n)$ is constructed to reflect increasing of the transitional barrier during transition due to internal and external components correlation:

$$F(\sigma_n) = 1 - \left(\frac{\sigma_n}{a}\right)^2. \tag{13}$$

The energy density function in the two-well form is the necessary condition of the existence of bistable conformational states in the system with energy (10,11):

$$E = \frac{1}{2}\left[\Phi(\sigma) + k\tau^2\right] - \chi F(\sigma), \tag{14}$$

where $\tau = R_n - R_{n-1}$ is the relative displacement of monomers. This condition gives restrictions on the value of $\chi$.

## STATIC EXCITATION OF THE SYSTEM

To consider static excitation of the system, let us change to a nondimensional variable $u_n = \sigma_n/a$. In the static case $\dot{u}_n = 0$, $\dot{R} = 0$ and thus total potential energy becomes:

$$E = \frac{1}{2}\sum_n \left\{\tilde{g}[u_n - u_{n-1}]^2 + k[R_n - R_{n-1}]^2 + \Phi(u_n) + \chi[R_{n+1} - R_{n-1}]F(r_n)\right\} \tag{15}$$

where $\tilde{g} = ga^2$.

To analyze possible excitations of the system (15) let us consider continual approximation:

$$E = \frac{1}{2h}\int_{-\infty}^{\infty} dz\left[\tilde{g}h^2 u'^2 + kh^2 R'^2 + \Phi(u) + 2\chi h R' F(u)\right] \tag{16}$$

where is the distance between monomers. The static equilibrium equations are:

$$R'' + \frac{\chi}{kh}\frac{\partial F(u)}{\partial u}u' = 0; \tag{17}$$

$$u'' - \frac{1}{2\tilde{g}h^2}\frac{\partial \tilde{\Phi}(u)}{\partial u} - \frac{\chi}{\tilde{g}h}R'\frac{\partial F(u)}{\partial u} = 0. \tag{18}$$

The system (17,18) has been solved with asymptotic of stable states:

$$u \to \pm 1, \ u'_z \to 0, \ R'_z \to 0, \ z \to \infty. \tag{19}$$

Therefore, static excitations have a form of two-component solitons.

$$u(z) = \pm th\left(\sqrt{\tilde{Q}}z\right); \tag{20}$$

$$R(z) = -\frac{\chi}{kh\sqrt{\tilde{Q}}}th\left(\sqrt{\tilde{Q}}z\right), \tag{21}$$

where $\tilde{Q} = \left(\varepsilon - \chi^2/k\right)/\tilde{g}h^2$.

The form of static deformation is described with a bell-shape soliton:

$$\rho(z) = hR' = -\frac{\chi}{k}ch^{-2}\left(\sqrt{\tilde{Q}}z\right). \tag{22}$$

The width of solitons is proportional to $L$:

$$L = \frac{1}{\sqrt{\widetilde{Q}}} = \sqrt{\frac{\widetilde{g}}{\varepsilon - \chi^2/k}}\, h\,. \qquad (23)$$

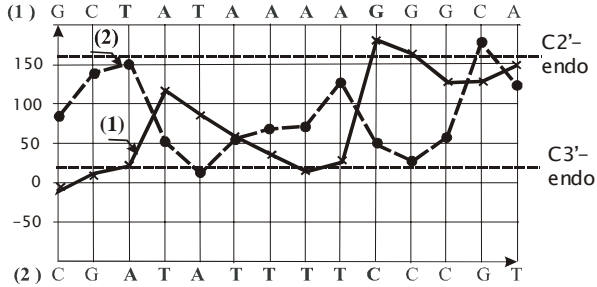

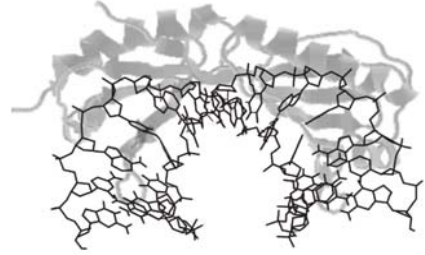

Fig.5: Sugar conformations (P°) for both chains of the TATA-box double helix [15]

Fig.6: Deformation of TATA-box bound to protein. Constructed according to (PDT024) [16] of data bank: http://ndbserver.rutgers.edu.

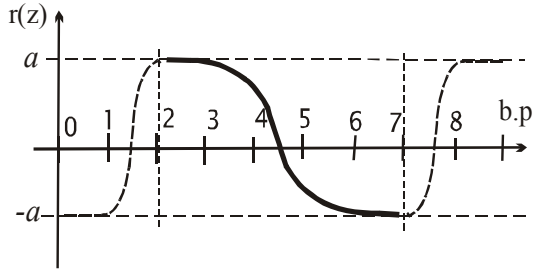

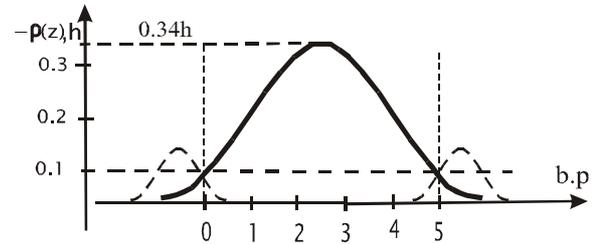

Fig.7: Conformational excitation of heteronomouse DNA (internal component).

Fig.8: Static deformation of heteronomouse DNA.

Thus conformational excitation in heteronomous DNA causes bend deformation of macromolecule with the form (22).

### ESTIMATIONS AND DISSCUSION

Conformation of heteronomous transition in the TATA-box of the DNA has been studied in the work [15] in detail. There are conformations of the sugars for both chains of double helix in this work. According to these data, the conformation of the sugars changes three times along the TATA-box and deformation of the box is realized in three bends (fig.5,6). Our results for the form of static excitations in heteronomous sequences is consistent with the form observed in the experimental work [15](fig.7,8).

In order to make quantitative comparison with experiment, we need to know a magnitude of static excitations $L$ in (20-22). According to the expression (23) this requires estimation of a correlation parameter and force constants of internal and external components $\chi$, $\widetilde{g}$, $k$. According to estimations [31], the force constants of internal and external subsystems are $\widetilde{g} = 40{,}08 \cdot 10^{-14}\, г$ and $k = 1519\, г/см^2$ ($\mu = 2{,}62 \cdot 10^{-37} г \cdot см^2$, $M = 9{,}84 \cdot 10^{-22} г$), respectively. The value of the parameter $\chi$ is chosen from the condition of a two-well form of the energy density function (14):

$$\chi < \sqrt{k\varepsilon}\,. \qquad (24)$$

Calculations for (24) give the value $\chi < 178{,}6 \cdot 10^{-7}\, г \cdot см/с^2$. For further calculations $\chi = 170 \cdot 10^{-7}\, г \cdot см/с^2$ has been used.

Using the obtained value of parameters, the width of a conformational soliton (22) can be calculated. The width of the soliton corresponding to the amplitude $0{,}1h$ (22) equals to approximately $5h$. A chosen amplitude approximately corresponds to amplitudes of DNA thermal oscillations.

A magnitude of deformation caused by conformational transition can be calculated with formula:

$$D = 2 \int_0^{5h/2} \rho(z) \frac{dz}{h} \approx 1{,}36 h. \qquad (25)$$

The obtained magnitude (25) corresponds to the bend of a chain approximately on $17°$.

Comparison of the parameters of the localized excitation with experimental data has been showed that such excitation can be realized in the TATA-box. According to developed theory, the experimental data obtained in S.K.Burley's and the co-authors work [15] can be explained by the fact of coexisting of three solitons: one is quite a wide in a center of the box and the rest are comparatively narrow solitons on the edges of the fragment. In a central part of box conformational excitation is conditioned directly by peculiarity of heteronomous form of the TATA-box. The width of a central excitation corresponds to the estimated width of static two-component soliton obtained in this study (fig. 7,8).

Two static localized excitations on the edges of TATA-box are caused by the close contact with protein complex loops. These excitations have the width of approximately 1h. Conformational solitons described in this paper can have this value of the width if the relation $\chi/kh$ is equal 0,137, which is approximately 2,5 times smaller than for the central soliton. It means that the interweaving of TBP protein loops between 1 and 2, and 7 and 8 monomers of TATA-box changes the parameters of the system. This problem must be considered separately. Preliminary estimation of deformation caused by either of these two solutions gives a bend of a chain on $7,5^o$. Therefore, according to our model, total deformation of TATA-box is approximately $32^o$. The estimated value of the bend is quite small comparatively with an experimental value (~$80^o$[13-15]). Such an essential difference is the result that we consider only the bend induced by conformational excitation heteronomuos DNA. Some part of deformation can be purely elastic. Also we might have overlooked some effects due to neglecting the deformation unwinding in the estimations. Thus, the model of heteronomous DNA transformations gives the result qualitatively describing a picture of TATA-box deformation. In further study of the quantitative description of transformations alternating sequences in DNA it is necessary to take into account the unwinding deformation, and also possible effects of system discontinuity.